# Electrical control of magnetism in oxides[*]

Cheng Song[†], Bin Cui, Jingjing Peng, Haijun Mao, and Feng Pan

Key Laboratory of Advanced Materials (MOE), School of Materials Science and Engineering, Tsinghua University, Beijing 100084, China

**Abstract**

This review article aims at illustrating the recent progresses in the electrical control of magnetism in oxides with profound physics and enormous potential applications. In the first part, we provide a comprehensive summary of the electrical control of magnetism in the classic multiferroic heterostructures and clarify their various mechanisms lying behind. The second part focuses on the novel route of electric double layer gating for driving a significantly electronic phase transition in magnetic oxides by a small voltage. The electric field applied on the ordinary dielectric oxide in the third part is used to control the magnetic phenomenon originated from the charge transfer and orbital reconstruction at the interface between dissimilar correlated oxides. At last, we analyze the challenges in electrical control of magnetism in oxides, both on mechanism and practical application, which would inspire more in-depth researches and advance the development in this field.

**Keywords:** electrical control of magnetism, oxide, magnetoelectric coupling, heterostructure

**PACS:** 75.47.Lx, 75.70.-i, 75.85.+t, 77.55.Nv

# 1. Introduction

[*] Project supported by the National Natural Science Foundation of China (Grant Nos. 51322101, 51202125 and 51231004) and National Hi-tech (R&D) project of China (Grant no. 2014AA032904 and 2014AA032901).
[†] E-mail: songcheng@mail.tsinghua.edu.cn

The electrical control of magnetism is an exciting new area of condensed-matter research with the potential to impacting the magnetic data storage, spintronics and high-frequency magnetic devices.[1,2] On paving its way towards spintronic-integrated circuits with ultralow power consumption, impressive improvements have been made to the properties of such nanoscale devices based on current-induced magnetization switching using spin-transfer torque (STT).[3,4] Further significant reductions in switching power are envisaged by using an electric field, which reduces the energy dissipation by a factor of 100 when compared with that in STT devices, making it comparable to that in the present semiconductor field-effect transistors, but with added non-volatile functionality.[1,2]

The electrical control of magnetism is of interest not only because of its technological importance, but also because it allows us to uncover properties of magnetic materials that are otherwise inaccessible.[2] In 2000, the electrical control of the magnetic phase transition was first demonstrated in magnetic semiconductor (In,Mn)As,[5] which was used as a channel material in a field-effect transistor. Following the investigations on magnetic semiconductors, researchers began to explore the electrical control of ferromagnetic metals. Despite that it was previously assumed that a large electric-field effect in metals would be difficult to observe[6] owing to the associated short screening length, the first observation of a direct electric-field effect on magnetism in a metal was reported for thin layers of FePt and FePd due to large electric-field-induced modulation of surface charges.[7] These successful studies encouraged more researchers to investigate the electrical control of magnetism in various systems, including the Fe(Co)/MgO[8,9] and CoFeB/MgO/CoFeB magnetic tunnel junctions (MTJs),[10,11] where the electrical control of magnetism is achieved by the change in interface magnetic anisotropy.

In the area of correlated complex oxide, the electrical control of magnetism has much broader playground[12], since the complex oxide which displays a variety of exotic properties such as ferromagnetic and superconductive two-dimension electronic gas, colossal magnetoresistance to be tuned by electrical method. Furthermore, owing to the insulating or semiconductor nature of complex oxide, they often has much longer screening length, ensuring an electric field of sufficient strength for magnetic phase control.[13] On the other hand, the more recent developments have benefited greatly from the tremendous progress achieved in the past two decades in the controlled epitaxial growth of complex oxides, which have allowed the fabrication of ultrathin oxide films and heterostructures with atomic precision grown in controlled environments; the progress in characterization tools increasingly dedicated to probing nanoscale phenomena; and the advances in first-principles calculations methods, which have enabled a deeper, atomic level understanding of the physical mechanisms underlying the strongly correlated behavior of metal oxides.[14] One of the exotic examples of electrical control of magnetism is in complex oxide multiferroics-based system, where the magnetoelectric (ME) coupling has enabled the magnetization switching induced by the application of electric fields to be enhanced.[2]

In terms of the mechanism involved in the electrical control of magnetism, there are a variety of answers. From the very beginning, the modulation of mobile carries or surface charge by electrical control is considered to play a dominant role especially in metallic magnetic materials[7,15] or diluted magnetic semiconductors (DMSs).[5,16] An induced or spontaneous electric polarization at the gate dielectric interface is screened by charge carriers from the channel layer, leading to charge accumulation or depletion and resulting magnetic properties. In the case of thin films of ferromagnetic semiconductors, this change in carrier density in turn affects the magnetic exchange interaction and magnetic anisotropy; in ferromagnetic metals, it instead changes the Fermi level position at the interface that governs the magnetic anisotropy of the metal.[2] Strain-mediated electrical control of magnetism is also a remarkable effect especially in piezoelectric-based or ferroelectric-based[17,18] or multiferroics-based systems.[19] The external electric field alters the lattice or shape of the ferroelectric crystal by the converse piezoelectric effect during switching, and then transfer the strain to the proximate magnetic layer, leading to the changes in magnetic anisotropy, magnetization rotation, and coercive through the magnetostriction. Another novel mechanism to be mentioned is the oxygen migration effect or redox reaction effect especially at metal-oxide interface or oxide heterostructures, where the migration of oxygen causes transition of ferromagnetic phase[20] or affects the orbital occupancy and magnetic anisotropy.[21]

The electric effect of magnetism is compatible with complementary metal–oxide–semiconductor (CMOS) technology and paves its way towards spintronic-integrated circuits with ultralow power consumption.[2] Our review is motivated to provide a comprehensive summary of the rapidly developing area of electrical control of magnetisms and clarify their various mechanism lying behind. While much remains to be explored in terms of materials optimization, development and characterization, our review is aimed at illustrating how novel functionalities can be generated by exploring the new phenomena arising at the interface between dissimilar materials and how such an approach can be used to achieve more efficient electrical control of magnetism.

## 2. Electrical control of magnetism in multiferroic heterostructures

Multiferroic materials with the coexistence of ferroelectric (breaking of space inversion symmetry) and (anti-)ferromagnetic (breaking of time reversal symmetry) orders have drawn ever-increasing interest due to their magnetoelectric (ME) coupling effect and potential for applications in multifunctional devices.[22,23] However the rare single-phase multiferroic materials exhibit ME coupling at low Curie temperatures, and a high ME coupling above room temperature has not yet been found in the single-phase systems, limiting their practical application.[24–26] $BiFeO_3$ (BFO) is unique with high Curie and Néel temperatures far above room temperature, but it is $G$-type antiferromagnetic or only very weak ferromagnetic.[23]

The development of ME coupling is fortunately lightened by the composite systems involving strong ferroelectric and ferromagnetic materials.[27,28] The greatly enhanced ME effect has been discovered experimentally in some two-phase nano-composites, such as $BaTiO_3$–$CoFe_2O_4$,[29] $PbTiO_3$–$CoFe_2O_4$,[30] and $BiFeO_3$–$CoFe_2O_4$.[31] The ME coefficient $\alpha$ in these composites is about three orders of magnitude higher than those aforementioned single phase ME materials. The similar mechanism has also been explored in the ferromagnetic/ferroelectric heterostructures, whose physical images are unambiguous and design flexibilities are greater. Next we focus on these multiferroic heterostructures.

The electric field control of magnetic behavior has been realized by the ME coupling effect in a great number of multiferroic heterostructures: the classic ferroelectric (FE) crystal and films serves as the ferroelectric layer while the performances of (anti-)ferromagnetic metals, oxides, and diluted magnetic semiconductors (DMS) are modulated.[14,32] A detailed summary of the ME coupling constants, results of electric fields, and corresponding mechanisms in different systems are provided in Table 1. One can see that the magnetic anisotropy, exchange bias, and magnetization, *etc.* are modulated with the effect of electric field [also shown in Figs. 1(a)–1(c)]. Following the extensive studies, the proposed mechanisms responsible for the ME coupling could be summarized as follows [Figs. 1(d)–1(f)]: i) the electrostriction introduces strain variation in the ferromagnetic (FM) layer, changing its lattice and concomitant magnetic properties; ii) the delicate modulation of carrier density by polarization reversal in FE field-effect transistor motivates the FM/antiferromagnetic (AFM) phase transition; iii) the use of multiferroic materials provides a route to the electrical control of spin arrangement by FM/AFM exchange coupling. Next we consider each of the different types of ME coupling in multiferroic heterostructures.

Table 1. Summary of electrical control of magnetism in common multiferroic heterostructures. $\alpha$ is given in the unit of Oe cm V$^{-1}$, $\alpha_\text{r}$ is the relative ME constant (dimensionless), and $T$ is the temperature. The abbreviations are defined as follows: La$_{1-x}$Sr$_x$MnO$_3$ (LSMO), La$_{1-x}$Ca$_x$MnO$_3$ (LCMO), SrRuO$_3$ (SRO), CoFe$_2$O$_4$ (CFO), BaTiO$_3$ (BTO), Pb(Mg$_{1/3}$Nb$_{2/3}$)O$_3$–PbTiO$_3$ (PMN-PT), BiFeO$_3$ (BFO), PbZr$_{1-x}$Ti$_x$O$_3$ (PZT), PbTiO$_3$ (PTO), YMnO$_3$ (YMO), exchange bias (EB), magnetic anisotropy (MA), magnetization ($M$), Curie temperature ($T_\text{C}$), coercivity ($H_\text{C}$).

| System | $10^3\alpha$ | $\alpha_r$ | $T$ (K) | Coupling | Results | Reference |
|---|---|---|---|---|---|---|
| LSMO ($x = 0.3$)/BTO | 20–230 | 6.0–69 | 200–300 | Strain | MA & $M$ | [17] |
| LSMO ($x = 0.33$)/PMN–PT | 60 | 18 | 300 | Strain | $T_C$ & $M$ | [18] |
| LCMO ($x = 0.3$)/PMN–PT | 34 | 10.2 | 10 | Strain | $M$ | [39] |
| CFO/PMN-PT | 7.5–9.6 | - | 300 | Strain | MA & $M$ | [41] |
| $Fe_3O_4$/CFO/PZT | 33 | 9.9 | 300 | Strain | $M$ | [44] |
| $Fe_3O_4$/PMN–PT | 108 | 32 | 300 | Strain | MA & $M$ | [45] |
| $Zn_{0.1}Fe_{2.9}O_4$/PMN–PT | 23 | 6.9 | 300 | Strain | MA | [46] |
| Ni/BTO | 0.5 | 0.15 | 300 | Strain | MA & $M$ | [49] |
| PZT/LSMO ($x = 0.2$) | 0.8–6.2 | 2.4–22 | 100 | Charge | $T_C$ & $M$ | [66,67] |
| LCMO ($x = 0.5$)/BFO | - | - | ≤300 | Charge | $M$ | [72] |
| SRO/ BTO | 1.1–5.9 | 0.32–1.8 | Theory | Charge | - | [73,74] |
| $Fe_3O_4$/ BTO | 20 | 5.7 | Theory | Charge | - | [75] |
| $CrO_2$/ BTO /Pt | 10 | 3.0 | Theory | Charge | $M$ | [76] |
| Co:$TiO_2$/PZT | - | - | ≤400 | Charge | $M$ & $H_C$ | [77] |
| Fe/ BTO | 3–16 | 0.9–4.8 | Theory | Charge | $M$ | [76,78,79] |
| Fe/ PTO | 73 | 22 | Theory | Charge | - | [79] |
| Ni/ BTO /Pt | 15 | 4.5 | Theory | Charge | $M$ | [76] |
| hcp Co/ BTO /Pt | 4 | 1.2 | Theory | Charge | $M$ | [76] |
| $Cr_2O_3$(111)/ [Co/Pt] | - | - | 150–250 | EB | EB, $H_C$ & $M$ | [92,93] |
| [Co/Pd]/$Cr_2O_3$(0001) | - | - | 303 | EB | EB | [94,95] |
| NiFe/YMO(0001) | - | - | 2–100 | EB | EB & $H_C$ | [96] |
| CoFe/BFO | - | - | 300 | EB & Strain | $M$ | [100] |
| LSMO ($x = 0.3$)/BFO | - | - | 5.5 | EB | EB & $H_C$ | [19,101] |

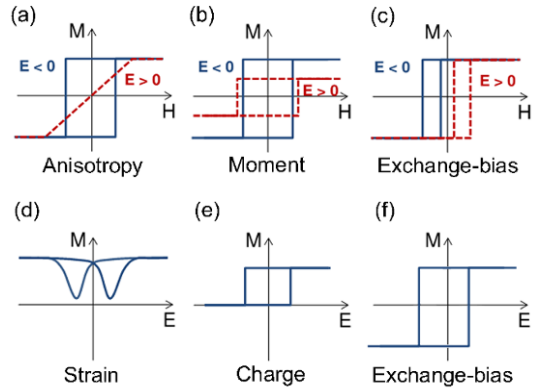

**Fig. 1.** (color online) Schematic of the different magnetic hysteresis curves expected in electrical control of magnetism (a)–(c) and the magnetoelectric coupling types in artificial multiferroics heterostructures (d)–(f).[14]

**2.1 Strain-mediated electrical control of magnetism**

Strain engineering is a powerful tool for manipulation of magnetism.[33–35] The strain-mediated electrical control of magnetism is considered to be the first attempt at designing a ME coupling in artificial FM/FE particulates and phase segregated ceramics,[36] and this approach has received renewed attention more recently.[37,38] This type of ME coupling were also widely reported in magnetic thin films (e.g. Ni, CoFeB, LSMO, and CFO) grown on a ferroelectric substrates (e.g. BTO, PMN-PT, and PZT).[17,18,39–58] The external electric field alters the lattice or shape of the ferroelectric crystal by the converse piezoelectric effect, and then transfers the strain to the proximate magnetic layer, leading to the changes in magnetic anisotropy, magnetization rotation, and coercivity through the magnetostriction. In the FM/FE heterostructures, both (anti-)ferromagnetic oxides and metals are controlled by electric field.

2.1.1 Ferromagnetic Oxides

In light of the rapid development of deposition technologies, such as oxide molecular beam epitaxy and reflection high-energy electron diffraction (RHEED)-assisted pulsed laser deposition (PLD), high-quality magnetic oxides could be epitaxially grown on the ferroelectric substrates, constructing the cornerstone of electrical control of magnetism mediated by strain.[59] In 2007, Eerenstein *et al.* found a giant change in the magnetic moment of epitaxial 40 nm LSMO ($x = 0.33$) on single crystal BTO substrate with the FE switching.[17] This variation in magnetization was attributed to the changes of magnetic anisotropy caused by strain coupling with non-180° BTO domain. The role of strain in this electrical manipulation was clearly demonstrated by Thiele *et al*. in the system of 20–50 nm LSMO ($x = 0.33$) and LCMO ($x = 0.33$) on PMN-PT substrate.[18] The dependence of magnetization on electric field follows the relationship between in-plane piezoelectric strain and electric field in a butterfly shape as shown in Fig. 2. However, this work explained the modulated magnetization to the paramagnetic-ferromagnetic phase transition driven by strain as the FE switching altered the Curie temperature.

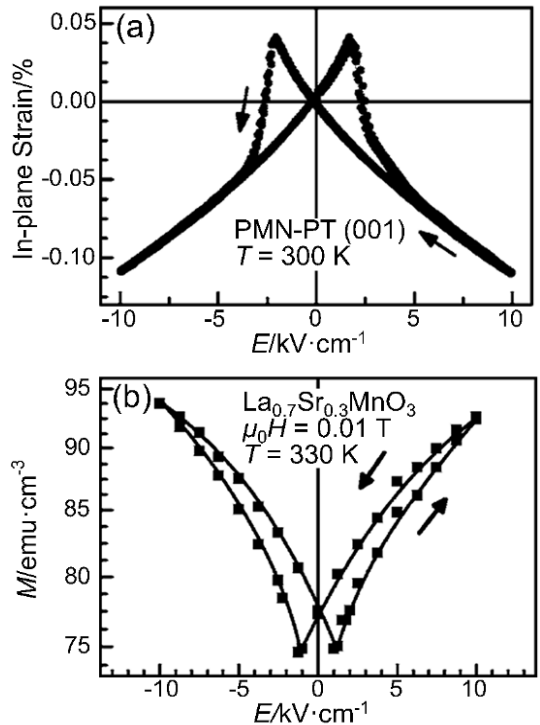

**Fig. 2.** (color online) (a) In-plane piezoelectric strain *vs* applied electric field ($E$ // [001]) recorded along a [100] edge of a 0.72PMN-0.28PT substrate. (b) Magnetization $M$ // [100] *vs* $E$ // [001] applied to the substrate for a LSMO/PMN-PT (001) heterostructure.[18]

Subsequently, strain-mediated electrical control of magnetism was observed in more and more oxides systems, with the realization of sharp and reversible changes in magnetization under electric field. However, the origins of strain-induced changes in magnetic properties have been vigorously pursued, with two mechanisms proposed to date: i) the electrical control of electronic phase transition and ii) the electrical modulation of magnetic anisotropy. The electronic phase transition driven by strain under electric field is supported by works using LSMO,[17,18] LCMO,[39] and $Pr_{0.6}Ca_{0.4}MnO_3$[40] with strong tendency toward phase separation as ferromagnetic layer. On the contrary, the changes of magnetic anisotropy under electric field was widely accepted in $CoFe_2O_4$[41,42] and $Fe_3O_4$,[43–45] where a magnetization orientation shift of 17 ° in $Fe_3O_4/BaTiO_3$ was observed.[43]

2.1.2 (anti-)ferromagnetic metals

Despite the well epitaxial oxide-based ferromagnet, magnetic metal thin films, which are usually polycrystalline or amorphous, can also be tuned by electric field. The strain transferred from the substrate is proved to be able to alter the magnetic anisotropy of magnetic metal films. Compared with the magnetic oxide, magnetic metal system attracts more attentions due to its high $T_C$, flexibility, and easy production. The FE control of magnetic metal was observed in many systems like Fe/BTO,[47] Co/PMN-PT,[48,49] Ni/BTO,[50] CoFe/PMN-PT,[51] CoPd/PZT,[52] Fe-Ga/BTO,[53] CoFeB/PMN-PT.[54,55] The magnetic anisotropy, coercivity, magnetic moment, and magnetic switching were controlled by electric field in these systems. In a Co/PMN-PT heterostructure, both simulations and experiments demonstrated a macroscopic maneuverable and non-volatile 180 ° magnetization reversal at room temperature as shown in Fig. 3(a).[49] The group of Zhao applied the electrical control of magnetization rotation in CoFeB/PMN-PT[54] into the CoFeB/AlO$_x$/CoFeB/PMN-PT magnetic tunnel junction, realizing a reversible, continuous magnetization rotation and manipulation of tunnel magnetoresistance (TMR) at room temperature by electric fields without the assistance of a magnetic field.[55] These works should be significant for the practical application of electrical control of magnetism in spintronics.

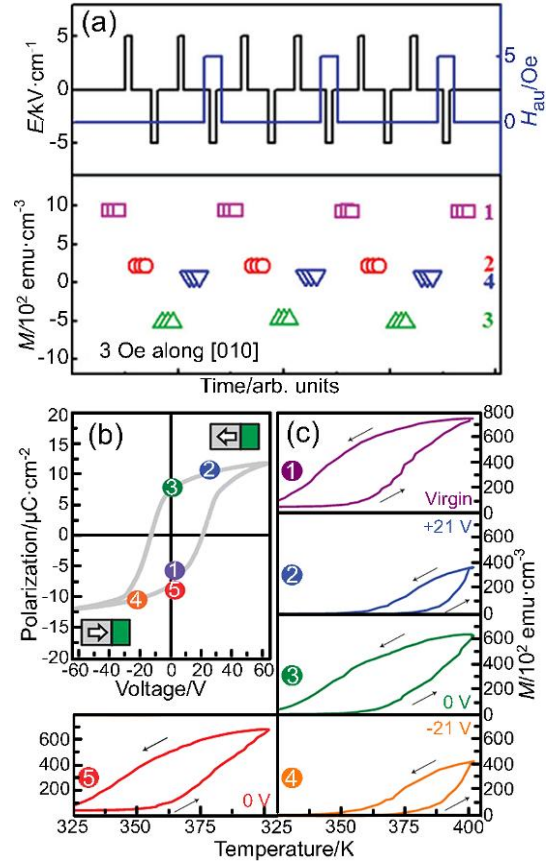

**Fig. 3.** (color online) (a) Pulsed electrical and magnetic operation and corresponding magnetic moments with a magnetic field of 3 Oe along [010] for Co/PMN-PT.[49] The different magnetization states are clarified as stage 1 ($M$ // [010]), stage 2 ($M$ // [$\bar{1}$00]), stage 3 ($M$ // [0$\bar{1}$0]), and stage 4 ($M$ // [$\bar{1}$00]).[56] (b) Polarization versus voltage loop collected at 300Hz and room temperature for BTO. (c) Temperature dependence of the magnetization of FeRh/BTO measured at 20 kOe for various voltages.

With the fast development of antiferromagnetic spintronics, more attempts are made in the control of antiferromagnetic materials by electrical means. Cherifi *et al* used ferroelectricity in BaTiO$_3$ crystals to tune the sharp metamagnetic transition temperature of epitaxially grown FeRh film and electrically drive a transition between AFM and FM order with only a few volts, just above room temperature [Figs. 3(b) and 3(c)].[56] Such a strain-mediated phase transition in FeRh under FE switching also produces a giant electroresistance response.[57]

**2.2 Charge-mediated electrical control of magnetism**

In heterostructures where the magnetic properties are intimately linked to charge, a change in carrier doping level would alter the magnetic performances. Two types of devices are developed to realize the charge-mediated electrical control of magnetism: ferroelectric field effect transistor and multiferroic tunnel junction.

2.2.1 Ferroelectric field effect transistor

The concept of field effect transistor (FET) could be taken as a reference here for designing artificial multiferroic structures to induce changes in the magnetic state. An induced or spontaneous electric polarization at the gate dielectric interface leads to charge accumulation or depletion and resulting magnetic property variations. For a ferroelectric material such as PZT, the charge carrier modulation is of the order of $10^{14}$ cm$^{-2}$, which is much larger than what is possible to attain using silicon oxide as the dielectric gate[60]. More importantly, this effect is non-volatile after removing the electric field. The ferroelectric field effect transistor (FE-FET) has been used to modulate a variety of properties including superconductivity[61,62] and metal–insulator transitions[63] for a long time. In the past decade, the control of magnetism by FE-FET was reported, both in first-principles calculations and experiments, for a variety of systems, such as complex oxides, DMS, and metal ferromagnets.

In 2009, the dramatic magnetoelectric coupling mediated by the modulation of hole-carrier density was predicted in La$_{1-x}$A$_x$MnO$_3$/BTO (001) ($A$ = Ca, Sr, or Ba) system by the first-principles method, where the doping concentration $x$ was 0.5, near the FM–AFM phase transition.[64] The direction of BTO polarization was used to electrostatically manipulate the hole-carrier density in La$_{0.5}$A$_{0.5}$MnO$_3$ and the resulting FM–AFM phase transition at the interface. When the polarization points away from or to the interface, there is an apparent upward or downward shift of the local density of states, favoring a hole charge accumulation (AFM) or depletion (FM) state, respectively. Besides, a microscopic model based on the two-orbital double-exchange was also introduced to describe this FE screening effect in manganites. The model simulation confirmed that the charge accumulation/depletion near the interface could drive the interfacial phase transition, which gave rise to a robust magnetoelectric response and bipolar resistive switching.[65]

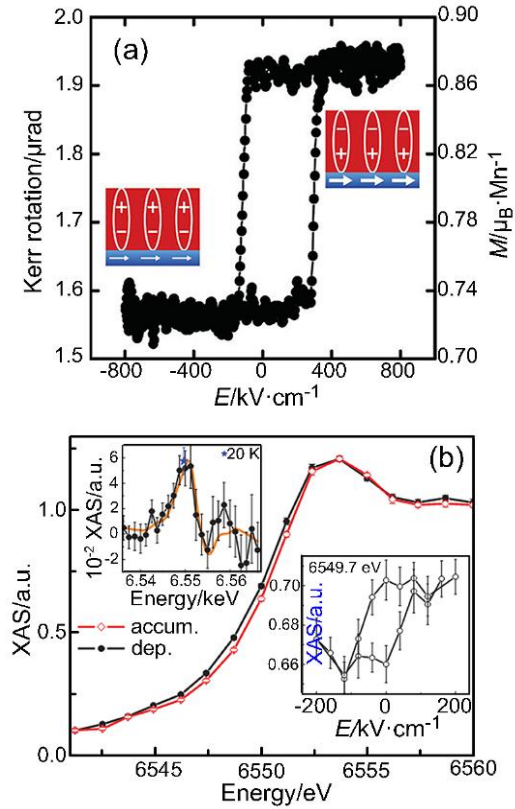

**Fig. 4.** (color online) (a) Magnetoelectric hysteresis curve of the PZT/LSMO at 100 K.[66] (b) Room temperature XANES results for the two polarization states of the PZT. The left top inset shows the difference in x-ray absorption for the two PZT polarization states. The right bottom inset is the variation of the x-ray light absorption as a function of the applied gate voltage at a fixed energy, $E = 6549.7$ eV.[67]

The electric field control of magnetism mediated by charge modulation in FM oxide films was firstly observed by Molegraaf *et al*.[66] The magnetic response of the PZT (250 nm)/LSMO (4 nm) as a function of the applied electric field exhibits in a magnetoelectric hysteresis curve as reflected by magneto-optic Kerr effect (MOKE) at 100 K in Fig. 4(a). The changes in carrier density are proved by x-ray absorption near edge spectroscopy (XANES) in Fig. 4(b)[67] and electron energy loss spectroscopy (EELS)[68]. Similar effect was observed or calculated in LSMO/BTO,[69–71] LCMO/BFO,[72] SRO/BTO,[73,74] $Fe_3O_4$/BTO,[75] $CrO_2$/ BTO /Pt,[76] PZT/Co:$TiO_2$,[77]

Fe/BTO,[78–80] and Co$_2$MnSi/BTO[81], *etc.*. The nature of the ME effect in these systems can be distinguished as follows: (i) the modification in the electronic structure like density of states near Fermi level[82] and electronic bonding[78,80] at the interface; (ii) changes in the magnetic exchange interaction;[77,83] (iii) changes in the magnetic anisotropy.[8,84] Sometime more than one of the above processes may occur simultaneously in a single system.

When we compare the strain- and charge-mediated cases, it is confusing that both the mechanisms of strain and carrier density appear in the same system like LSMO/BTO[17,69,70] and Fe/BTO.[47,78–80] It is noteworthy that the modulation of carrier density induced by the electric field or polarization is mainly concentrated in the area near the interface, whose thickness is determined by the screening thickness ($t_S$) of the FM layer:[6]

$$t_S \approx (\varepsilon \hbar^2 / 4me^2)^{1/2} (1/n)^{1/6} \tag{1}$$

where $\varepsilon$ is the dielectric constant, $\hbar$ is the Planck constant, $m$ is the electron mass, $e$ is the electron charge, and $n$ is the carrier density. The $t_S$ value is around 2 nm for commonly metallic FM material, resulting in a very thin FM layer in the charge-mediated cases. In contrast, the thickness of FM layer is not limited by the $t_S$ value, which could be larger than 10 nm sometime in the strain-mediated case. Moreover, the thick FE substrate in strain-mediated case could produce a significant strain on the FM layer above, while the reduced FE thin film thickness in charge-mediated case might weakens the clamping effect of FE-induced strain on the FM layer below. Additionally, the strain effect mainly comes from the lattice distortion between polarized and unpolarized states for the common (001) BTO and PMN-PT ferroelectric layer, but the polarization switching from positive to negative in charge-mediated case has little change in lattices.[30,50] Hence, the electrical control of magnetism mediated by strain follows the dependence of strain on electric field while that mediated by carrier density follows the dependence of polarization on electric field.

2.2.2 Multiferroic tunnel junction

The electrical control of magnetism can also been realized in multiferroic tunnel junctions (MFTJs) where a FE (multiferroic) tunnel barrier is sandwiched between two FM (a FM and a normal metallic) electrodes.[85−91] The interaction between ferroelectricity and ferromagnetism can be studied through transport measurement, which displays four distinct resistance states due to the TMR and tunnel electroresistance (TER) effects as shown in Fig. 5(a). The four resistance states were first reported by Gajek *et al.* in a LSMO/La$_{0.1}$Bi$_{0.9}$MnO$_3$ (LBMO)/Au MFTJ [Fig. 5(b)],[87] but the obvious variation of the TMR with the ferroelectric polarization have not been observed, and the multiferroic materials are very rare. Subsequently, artificial MFTJs with FE barrier and FM electrodes were designed. Garcia *et al.* demonstrated that the TMR of −19% and −45% could be detected in LSMO ($x = 0.33$)/BTO (1 nm)/Fe MFTJ with FE layer under downwards and upwards states, respectively [Fig. 5(c)].[88] In the work of Pantel *et al.* even the sign of the TMR was reversed in the LSMO ($x = 0.33$)/PZT (3.2 nm)/Fe MFTJ by the ferroelectric polarization, the TMR values were +4% and −3% when the polarizations were pointing to LSMO ($x = 0.33$) and Co, respectively [Fig. 5(d)].[89] Moreover, Yin *et al.* showed that the inserting of 0.8 nm interfacial LCMO ($x = 0.5$) in the LSMO ($x = 0.3$)/LCMO ($x = 0.5$)/BTO/LSMO ($x = 0.3$) MFTJ gave rise to a 500% variation of the TMR owing to the FM-AFM phase transition induced by the polarization-generated metal/insulator phase transition in LCMO ($x = 0.5$) [Fig. 5(e)].[90] Such a resistive switching and phase transition with the hole modulation driven by FE polarization has also been predicted by the theoretical work from Tsymbal's group.[91] Unfortunately, all the aforementioned works of ferroelectric control of spin-polarization can be only performed below room temperature, and further efforts should been devoted to obtain the efficient ferroelectric control of spin-polarization at room temperature.

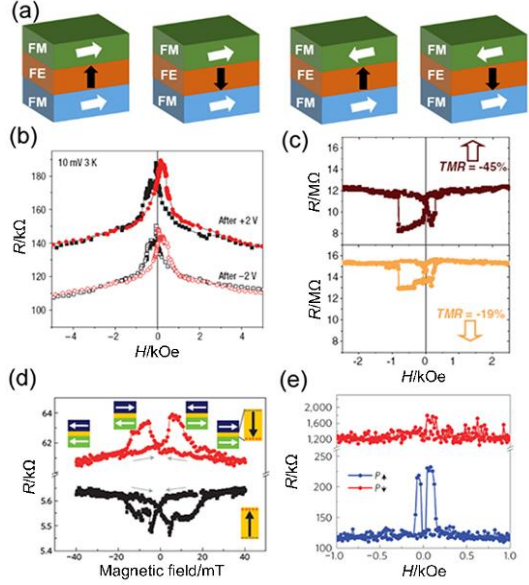

**Fig. 5.** (color online) (a) Schematic of the four resistance states in artificial MFTJS with two FM electrodes sandwiching a FE tunnel barrier. The white and black arrows represent the magnetic configurations and FE polarization, respectively. (b) Tunnel magnetoresistance curves at 4 K with $V_{dc.}$ = 10 mV in an LSMO ($x$ = 0.33)/LBMO(2 nm)/Au junction.[87] (c) Resistance versus magnetic field ($R$–$H$) curves for LSMO ($x$ = 0.33)/BTO (1 nm)/Fe MFTJ ($V_{dc.}$ = −50 mV, $T$ = 4.2 K) after poling the ferroelectric barrier up or down, respectively.[88] (d) $R$–$H$ curves measured at 50 K in the as-grown state of LSMO ($x$ = 0.33)/PZT (3.2 nm)/Fe MFTJ.[89] (e) $R$–$H$ curves for LSMO ($x$ = 0.3)/ LCMO ($x$ = 0.5)/BTO/ LSMO ($x$ = 0.3) MFTJ.[90]

**2.3 Exchange bias-mediated electrical control of magnetism**

The exchange bias effect is known to be associated with the coupling between FM and AFM materials, where the exchange coupling gives rise to a shift of the magnetic hysteresis loop away from the center of symmetry at zero magnetic field. When the single phase multiferroic materials like $Cr_2O_3$,[92–95] $YMnO_3$,[96,97], $LuMnO_3$[98], and $BiFeO_3$[19,99–102] (FE and AFM orders) serve as the FE layer, the exchange bias of heterostructure might be electrically reversed. The manipulation of exchange bias using intrinsic multiferroics allows the possibility of electric field switching of the direction of the magnetization.

Demonstration of exchange bias controlled by electric fields was first reported in perpendicularly magnetized [Co/(Pt or Pd)]/$Cr_2O_3$ heterostructures.[92–95] Reversible switching of the exchange bias was observed at 303 K in [Co/Pd]/$Cr_2O_3$(0001) heterostructure by the simultaneous application of a magnetic and electric field to switch the antiferromagnetic single domain state, which in turn switched the direction of the uncompensated spins at the $Cr_2O_3$ (0001) interface that biased the [Co/Pd] hysteresis loop.[92] The electric field control of exchange bias in NiFe/$YMnO_3$ was reported by Laukhin *et al*,[96] who showed that the exchange bias after cooling the system under a magnetic field could be reduced to near zero by applying a voltage across the $YMnO_3$.[97]

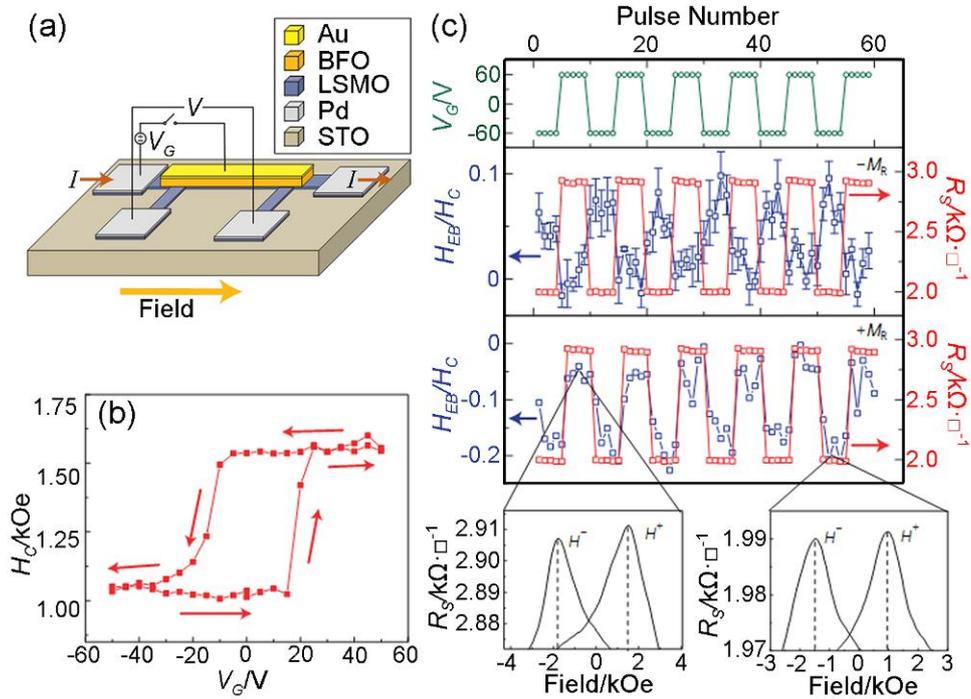

**Fig. 6.** (color online) (a) A schematic of the BFO/LSMO field-effect device. (b) Magnetic coercivity of the LSMO with respect to gate voltage ($V_G$) at 5.5 K. The arrows show the direction of the pulse sequence. (c) Electric-field control of exchange bias. From top to bottom: the $V_G$-pulse sequence used for the measurements; the measurements of normalized exchange bias and peak resistance for the gate-pulse sequence obtained in negative and positive remanent magnetization; examples of individual MR curves from the upper and lower resistive states where the exchange-bias values were determined.[19]

BFO is the most famous multiferroic materials with a ferroelectric $T_C$ of ~1100 K and *G*-type AFM Néel temperature of ~640 K.[24,103–106] The origin of exchange bias in *G*-type AFM BFO, with a fully compensated spin configuration at the interface, is related to the Dzyaloshinskii-Moriya interaction and FE polarization, which was theoretically demonstrated in the systems of BFO/LaMnO$_3$[107] and SrMnO$_3$/SrRuO$_3$ by Dong *et al*.[108] The FE polarization of BFO can point along any one of the eight degenerate [111] directions, permitting the possible formation of 3 types of domain walls (71°, 109°, and 180°).[99,104] The domain structures could be controlled by the growth orientation, strain, and substrate.[109–111] In 2006, Zhao *et al* demonstrated the first observation of electrical control of antiferromagnetic domain structure in BFO at room temperature, where the FE polarization switching would induce the AFM domain switching.[99] Subsequently, the exchange bias in a field-effect device with 600 nm BFO as the dielectric and 3 nm LSMO ($x$ = 0.3) as the conducting channel can reversibly switch between two distinct states by poling the ferroelectric polarization of BFO in Fig. 6.[19] Although the role of strain and charge could not be fully excluded here, Wu *et al* thought that the electrical manipulation of exchange bias mainly arose from the induced interfacial magnetism in a few nanometers of the BFO layer. The interfacial magnetism in BFO was attributed to an electronic orbital reconstruction occurring at the BFO/LSMO interface.[101,112]

Although multiferroic materials, e.g. BFO, are the typical FE layer for exchange bias-mediated electrical control of magnetism, it is inaccurate to consider that all the electrical control of magnetism in the systems including BFO originates from exchange bias. For instance, an electrically driven change in FE polarization and thus AFM order in BFO could switch the magnetization of thin metal film (e.g. CoFe) above.[100] The mechanical strain produced by ferroelastic could also modify the preferred orientations of the magnetic domains and therefore the macroscopic magnetization. When a material near FM-AFM phase transition point, like LCMO ($x$ = 0.5), was combined with BFO, the modulation of carrier caused by ferroelectric polarization might play a dominated role in the electrical manipulation of magnetism with competition between FM and AFM phases.[72]

## 3. Electrical control of magnetism by ionic liquid

Field-effect transistors are indispensable for information process and widely used in electrical control of magnetism. In the FET, the capacitance ($C$) of insulating layer determines the change in carrier density $\Delta n_S$ ($\Delta n_S \sim CV_G/e$, $V_G$ is gate voltage and $e$ is the electron charge). To realize a larger $\Delta n_S$ without increasing $V_G$, the enhancement of $C = \kappa\varepsilon_0/d$ ($\kappa$ is the relative permittivity, $\varepsilon_0$ is the vacuum permittivity, and $d$ is the thickness of the dielectric layer) is an effective route.[113] A conventional approach to obtaining a large $C$ is to adopt high $\kappa$ materials (e.g. $Al_2O_3$, $HfO_2$, or $ZrO_2$) with a thickness of ~50 nm by atomic layer deposition. Alternatively, ionic liquid consisting of anions and cations could produce an electric double layer (EDL) (pairs of sheets of negative and positive charges) at the interface between ionic liquid and channel under electric field. The EDL behaves a very large $C$ because the gap between the two charged sheets, which corresponds to $d$, is on the order of the size of the anions and cations as shown in Fig. 7. And thus a dramatically large $\Delta n_S$ is realized. The common cations in ionic liquid include N,N-diethyl-N-(2-methoxyethyl)-N- methylammonium (DEME), 1-ethyl-3-methylimidazolium (EMIM), N-methyl-N- propylpiperidinium (MPPR), 1,3-diallylimidazolium (AAIM), 1-al-lyl-3- ethylimidazolium (AEIM), 1-allyl-3-butylimidazolium (ABIM), N,N,N- trimethyl-N- propylammonium (TPA), $KClO_4$, and $CsClO_4$, *etc.*, while the common cations include bis-(trifluoromethylsulfonyl)imide (TFSI), polyethylene oxide (PEO), and $BF_4$, *etc.*.

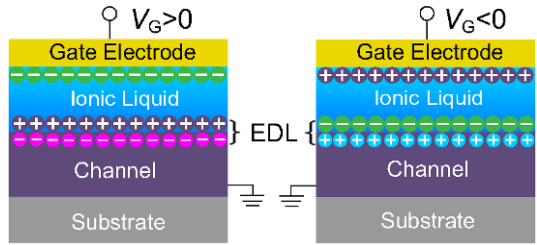

**Fig. 7.** (color online) Schematic illustration of the device structure of EDL under positive (left) and negative (right) gate voltage.

The EDL gating is firstly used in the electrical manipulations of superconductivity, two-dimensional conductivity, and metal-insulator transition. Ueno *et al.* found the electric-field-induced superconductivity in the insulating SrTiO$_3$ by using the EDL in KClO$_4$-PEO.[114] The sheet carrier density was enhanced from zero to $10^{14}$ cm$^{-2}$ and a two-dimensional superconducting state emerged below a critical temperature of 0.4 K under a gate voltage ($V_G$) of 3.5 V. The EDL also brought about significant modulations of carrier density and sheet resistance in the two-dimensional conductive system such as LaAlO$_3$/SrTiO$_3$,[115] graphene,[116] and MoS$_2$.[117] Nakano *et al.* found that a $V_G$ applied on the vanadium dioxide (VO$_2$) through DEME-TFSI drove all the previously localized charge carriers in the bulk material into motion, leading to the emergence of a three-dimensional metallic ground state.[118] In a voltage-sweep measurement, the metal–insulator transition provided a non-volatile memory effect, which was operable at room temperature. Besides, the metal-insulator transitions in other systems like SmCoO$_3$,[119] NdNiO$_3$,[120] and Ca$_{1-x}$Ce$_x$MnO$_3$,[121] *etc.* were also realized by electrical means using ionic liquid. More recently, the magnetism in both metal and oxide system is electrically tuned by EDL gating. For instance, Weisheit *et al.* and Wang *et al.* manipulated the coercivity in FM FePt (FePd) and exchange-spring in AFM IrMn, utilizing the EDL gating, respectively.[7,122] However, the mechanism for EDL gating is under intense debate: electrostatic doping and electrochemical reaction. In the following, we will discuss the electrical control of magnetism in oxides with different mechanisms.

**3.1 Electrostatic doping**

According to the conventional understanding from capacitor model, the electric field only changes the carrier density ($\Delta n_S \sim CV_G/e$) in the system by an electrostatic doping mechanism. Dhoot *et al.* studied electrostatic field-induced doping in LCMO ($x = 0.2$) transistors using electrolyte (EMIM-TFSI) as a gate dielectric as shown in Fig. 8 (a). For positive gate bias, electron doping drives a transition from a ferromagnetic metal to an insulating ground state.[123] The thickness of the electrostatically doped layer depended on bias voltage but could extend to 5 nm requiring a field doping of $2 \times 10^{15}$ charges per cm$^2$ equivalent to 2.5 electrons per unit cell area. In contrast, negative gate voltage enhanced the metallic conductivity by 30%. In $Pr_{1-x}Sr_xMnO_3$ ($x = 0.5$) thin film, the doping level $x$ was controlled by gate voltage, accompanied by the modulation of Curie temperature and resistance, realizing a phase transition similar to the classic phase diagram.[124]

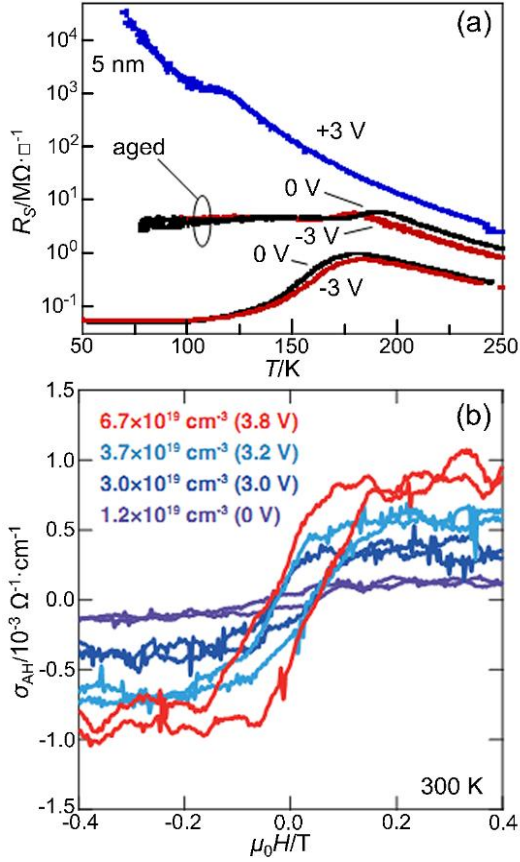

**Fig. 8.** (color online) (a) Sheet resistance versus temperature for 5 nm LCMO ($x = 0.2$) at $V_G = 0$ (black line), $-3$ (red line), and $+3$ V (blue line).[123] (b) Magnetic field dependence of the anomalous Hall conductivity $\sigma_{AH}$ at 300 K for Co: $TiO_2$, measured at different $V_G$. The values of electron density n at each $V_G$ obtained from the ordinary Hall effect are shown in parentheses.[125]

In other system like magnetic oxide semiconductor of Co: $TiO_2$, Yamada *et al.* realized electric field–induced ferromagnetism at room temperature by means of EDL gating with high-density electron accumulation ($>10^{14}$ per square centimeter).[125] By applying a gate voltage of a few volts, a low-carrier paramagnetic state was transformed into a high-carrier ferromagnetic state, thereby revealing the considerable role of electron carriers in high-temperature ferromagnetism and demonstrating a route to room-temperature semiconductor spintronics as shown in Fig. 8(b). The carrier density, resistance, and magnetoresistance were also tuned by EDL gating in the 5$d$ iridates of $Sr_2IrO_4$ with large spin-orbit coupling.[126]

**3.2 Electrochemical reaction**

Although the electrostatic doping mechanism successfully figures out the changes in carrier density in some systems, it cannot explain the non-volatile electric-field effect in the whole bulk of film. Thus another mechanism base on the electrochemical reaction is then proposed: the electric field applied on ionic liquid could bring about redox in the channel material, which could induce the creation and migration of oxygen vacancies.[127] The introduction of oxygen vacancies makes the non-volatile and in-depth effect reasonable.

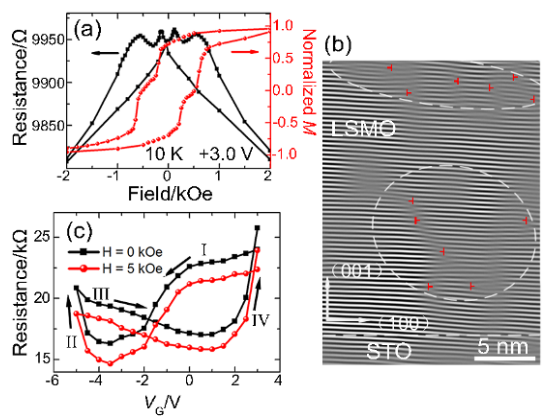

**Fig. 9.** (color online) (a) Channel magnetoresistance acquired by sweeping the magnetic field along the channel (left axis) and the normalized magnetization curves measured with the magnetic field applied in-plane along the (100) direction of the substrate at 10 K under $V_G$ = +3 V. (b) Fourier-filtered images of LSMO with $V_G$ = +3.0 V. The ⊥ and dashed ovals mark the dislocation and estimated areas with high dislocation density, respectively. (c) Channel resistance versus $V_G$ with $H$ = 0 kOe and 5 kOe.[20]

For example, the electrically reversible control of ferromagnetic phase transition based on oxygen vacancies migration in manganite film was reported by Cui *et al.*.[20] The formation of an insulating and magnetically hard phase induced by the migration of oxygen in the magnetically soft matrix was directly observed in the magnetoresistance [Fig. 9(a)] and Fourier-filtered images of [Fig. 9(b)], which randomly nucleated and grew across the film instead of initiating at the surface and spreading to the bottom. This discovery provides a conceptually novel vision for the electric-field tuning of phase transition in correlated oxides. The realization of a reversible metal-insulator transition in colossal magnetoresistance materials will also further the development of four-state memories, which can be manipulated by a combination of electrode gating and the application of a magnetic field [Fig. 9(c)]. The reversible control of charge transport, metal-insulator crossover and magnetism in field-effect devices were also realized based on EDL gated SRO.[128] In these thin-film devices, the metal-insulator crossover temperature and the onset of magnetoresistance could be continuously and reversibly tuned in the range 90–250 K and 70–100 K, respectively, by application of a small gate voltage. A reversible diffusion of oxygen ions in the oxide lattice dominates the response of these materials to the gate electric field.

Electrical manipulation of lattice, charge, and spin is realized respectively by the piezoelectric effect, field-effect transistor, and electric field control of ferromagnetism, bringing about dramatic promotions both in fundamental research and industrial production. However, it was generally accepted that the orbital of materials were impossible to be altered once they have been made. The EDL gating was used to dynamically tune the orbital occupancy and corresponding magnetic anisotropy of LSMO thin films in a reversible and quantitative manner.[21] Positive gate voltage increases the proportion of occupancy of the orbital and magnetic anisotropy that were initially favored by strain (irrespective of tensile and compressive), while negative gate voltage reduces the concomitant preferential orbital occupancy and magnetic anisotropy (Fig. 10).

The electrochemical reaction mechanism was also demonstrated in the Co/GdO$_x$, where the redox at the interface was controlled by the migration of oxygen ions under electric field.[129,130] The Co films could be reversibly changed from an optimally oxidized state with a strong perpendicular magnetic anisotropy to a metallic state with an in-plane magnetic anisotropy or to an oxidized state with nearly zero magnetization, depending on the polarity and time duration of the applied electric fields. Yuan *et al*. figured out a "phase diagram" to distinguish the electrostatic or electrochemical nature of EDL gating: the high work frequency and low temperature favors the electrostatic doping, while low frequency and high temperature favors the electrochemical reaction.[131] Nevertheless, the factors that determine the mechanism of EDL gating are various, e.g. strain state, and oxygen diffusion ability, *etc*, which need more in-depth investigations.

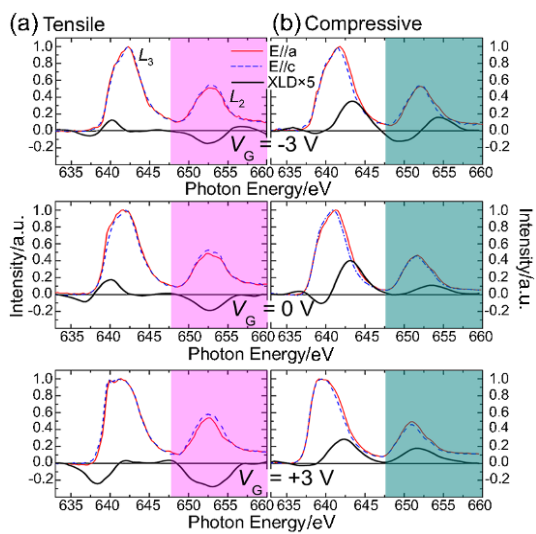

**Fig. 10.** (color online) Normalized X-ray absorption spectroscopy (XAS) [photon polarization parallel (E//a) and almost perpendicular (E//c) to the sample plane] and x-ray linear dichroism (XLD) signals of (a) tensile and (b) compressive strained LSMO.[21]

## 4. Electrical control of magnetism by ordinary dielectric oxide
### 4.1 LAO/STO interface

The two-dimensional electron gas (2-DEG) or two-dimensional electron liquid (2DEL) that forms at the interface between the two insulating non-magnetic oxides $LaAlO_3$ (LAO) and $SrTiO_3$ (STO) has sparked widespread research interest due to its possession of a remarkable variety of emergent behavior including superconductivity,[132,133] strong Rashba-like spin-orbit coupling[134] and ferromagnetism.[135-138] From a fundamental perspective, interfacial ferromagnetism could be a leading example of truly emergent phenomena; although bulk $SrTiO_3$ can be doped to be metallic and superconducting, neither constituent in bulk form exhibits ferromagnetism.[138] The first signatures of magnetism at the LAO/STO interface were reported in magnetotransport measurements by Brinkman *et al*.[139] Direct current scanning quantum interference device (SQUID) magnetometry measurements by Ariando *et al*.[140] showed ferromagnetic hysteresis extending to room temperature. Torque magnetometry measurements by Li *et al*.[136] showed evidence for in-plane magnetism with a high moment density. Scanning SQUID microscopy by Bert *et al*.[137] revealed inhomogeneous micron-scale magnetic 'patches'. X-ray circular dichroism measurements by Lee *et al*.[138] indicated that the ferromagnetism is intrinsic and linked to $d_{xy}$ orbitals in the Ti $t_{2g}$ band. Despite this variety of evidence, the existence and nature of magnetism in LAO/STO heterostructures has remained controversial. Since most of the exotic properties depend strongly on carrier density that can be tuned easily by electric field,[134] the nature of 2-DEG or 2-DEL can be readily explored by an electric field including back-gating,[141] top-gating,[142] or via nanoscale control using conductive atomic force microscopy (AFM) lithography.[143]

At the LAO/STO interface, the electron gas is inherently sandwiched between two insulators. It is thus rather natural to try and explore the system's ground states by modulating the carrier density with an electric field. In a standard field-effect device, an electric field is applied between a metallic gate and a conducting channel across a dielectric. SrTiO$_3$ substrate is chosen as the dielectric because it is characterized at low temperatures by a large dielectric constant. The metallic gate is a gold film sputtered opposite to the channel area onto the back of the substrate. A typical sketch of the back-gated field-effect device is shown in the inset of Fig. 11(a). A series of experiments, using the STO substrate as the gate dielectric, has revealed a complex phase diagram [shown in Fig. 11(a)].[134,144] A superconducting dome appears within a certain range of carrier densities modified by electric field. Reducing the carrier concentration from the largest doping level ($V$~320 V), critical temperature $T_{BKT}$ first increases, reaches a maximum at around 310 mK and then decreases to zero. This critical line ends at $V_C < -140$ V, where the system undergoes a quantum phase transition. Transport measurements have proved the existence of a very large perpendicular magnetoresistance (MR), interpreted as evidence of weak (anti)localization effects. Close to $V_C$, the magnetoconductance switches from positive to negative, as displayed in Fig. 11(b). It was believed that this evolution is tied to the steep rise of a Rashba spin-orbit interaction upon increasing the gate voltage.[144]

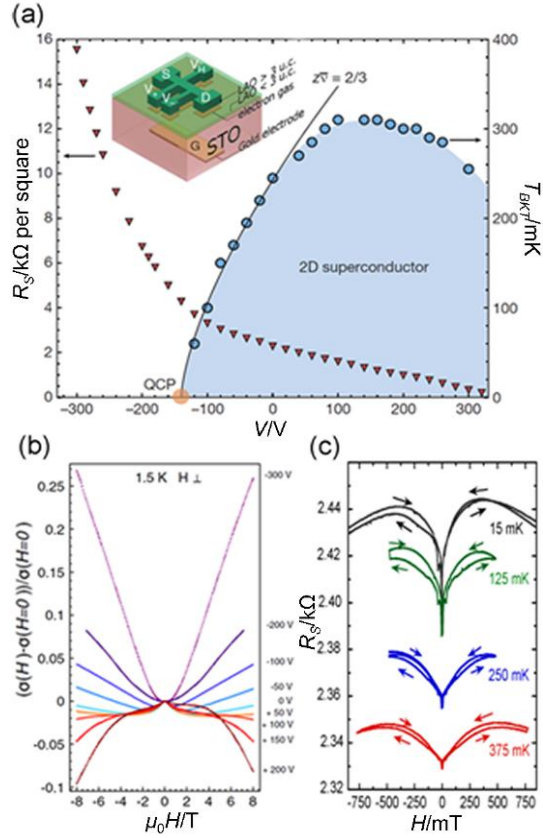

**Fig. 11.** (Color online) (a) Electronic phase diagram of the LAO/STO interface obtained through field-effect modulation of the carrier density.[144] The inset show schematic view of a field-effect device, showing the source (S), drain (D), longitudinal voltage ($V_+$ and $V_-$), Hall voltage ($V_H$) and gate voltage (G) contacts. (b) Modulation of the magnetoconductance Δs with gate voltage for a LAO/STO interface.[134] The change from a weakly localized regime (positive Δs) to a weakly anti-localized regime (negative Δs) reveals the increase in strength of the spin-orbit coupling. (c) Magnetoresistance (MR) at $V_g$~100 V at different temperatures.[145]

Since the electronic properties of LAO/STO interfaces are extremely sensitive to growth conditions, some other sample prepared under higher oxygen pressure display coexistence of superconductivity and ferromagnetism.[145] Figure 11(c) shows the MR of the sample at a few different temperatures at $V_g \sim 100$ V (not in the superconducting regime). The magnitude of the resistance change and the sharpness of the resistance dip near zero field increase with decreasing temperature, consistent with a phase coherence length that increases with decreasing temperature. A closer look at the low-field MR reveals an additional resistance dip near zero field that is related to weak localization. The MR also shows a hysteretic butterfly pattern similar to that seen in $T_c(H)$ (not shown here), indicating that local magnetic fields arising from magnetic order also modulate quantum interference in the carrier gases.

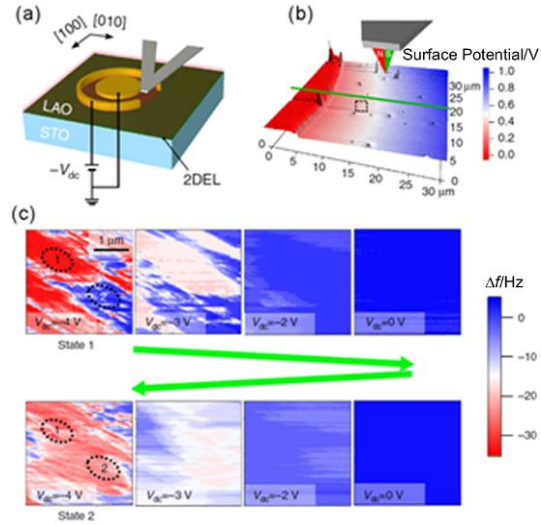

**Fig. 12.** (Color online) (a) Experimental set-up, The MFM tip is mechanically driven by a piezoelectric transducer near its resonant frequency and kept at a constant height above the surface. The top electrode and MFM tip are both grounded and a DC bias is directly applied to the interface.[135] (b) Kelvin-probe force microscopy measurement of a region that includes the area for which MFM measurements are made.[134] The top gate is grounded and voltage bias to the interface ($V_{dc}$) is 3 V. The topography is shown as height, while the colour maps onto the measured surface potential (the work function is already subtracted). (c) MFM frequency images over a 3×3 mm area indicated by the black dashed line enclosed region in b. The MFM tip is magnetized horizontally parallel to the [010] sample direction.[135] MFM frequency images for $V_{dc}$ increasing from −4 to 0 V then decreasing to −4 V. Magnetic domain features are clearly observed for $V_{dc} < −2$ V. The final State 2, obtained after cycling the voltage to $V_{dc} = 0$, is uncorrelated from the initial State 1. The regions enclosed by dashed lines

give example of where magnetic contrast is unchanged (region 1) or reversed (region 2) after voltage cycling.

Aside from back gating, there is also top-gating of 2DEG by employing the dielectric of LAO film, which is often as thin as several unit cell and needs much lower gating voltage. A two-terminal capacitor device is used to electrically gate the LAO/STO interface [Figs. 12(a) and 12(b)]. The top circular electrode is grounded and a voltage $V_{dc}$ is applied to the annular interface contact. Decreasing $V_{dc}$ depletes the interface of mobile electrons, while increasing $V_{dc}$ leads to electron accumulation and results in a conductive interface. By using top gating LAO/STO interface with different voltages, interfacial magnetism as a function of mobile interfacial carrier density can be detected by magnetic force microscopy (MFM). As shown in Fig. 12(c), the first MFM image, taken at $V_{dc}$~4 V ('State 1'), shows strong contrast electrode in the frequency channel, signaling clear out-of-plane ferromagnetic domain. As the voltage goes to –2 V, the domain contrast has nearly vanished with new horizontal bands appearing parallel to the fast scan axis until the contrast is absent at 0 V. The phenomena clearly indicate the emergence of an in-plane ferromagnetic phase as electrons are depleted from the interface.

The discovery of electrically controlled ferromagnetism at the LAO/STO interface provides a new and surprising route to a wide range of spintronics applications. Nevertheless, the nature of the ferromagnetic state at the intriguing LAO/STO interface remains unexplored, and many effects such as spin-torque transfer, spin-polarized transport, and electrically controlled spinwave propagation and detection are expected but not demonstrated. Thus, there are a plenty of space to employ the electric control in controlling interfacial ferromagnetism for the versatile application of the intriguing LAO/STO interface.

**4.2 Other dielectric oxides**

Controlling magnetization switching by an electric field without the assistance of a bias magnetic field may enable new data-storage spintronic devices acquiring low electric power. The modification of the correlated oxide interface may offer an avenue to explore this new strategy. In the experiments by Cuellar *et al.*,[146] the manipulation of interfacial magnetic Cu states with electric field in $La_{0.7}Ca_{0.3}MnO_3$/$PrBa_2Cu_3O_7$/$La_{0.7}Ca_{0.3}MnO_3$ magnetic tunnel junctions [Fig. 13(a)] realized the electrical switching of the magnetization without an applied magnetic field. As shown in Fig. 13(b) both high and low resistance states corresponding to two different magnetic configuration are stable in zero magnetic field, in other words using electrical means only can stabilize two different magnetic states at very low power.

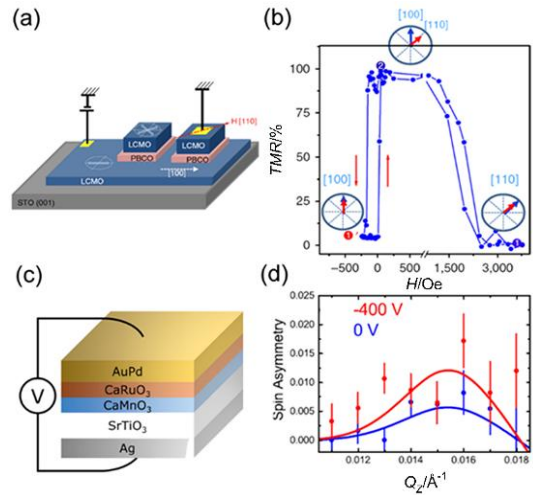

**Fig. 13.** (color online) (a) The sketch of the LCMO ($x = 0.3$)/$PrBa_2Cu_3O_7$/LCMO ($x = 0.3$) magnetic tunnel junctions and the test setup. Magnetic field was directed in-plane along the [110] direction (easy axis of the harder top layer), according to drawing.[146] (b) At low bias (100mV, blue symbols) the minor loop shows the saturation State 1, the high resistance State 2 due to misalignment of magnetic moments of top and bottom layer, and the low resistance 1´ where magnetic moments are aligned along the easy axis of the bottom layer. Notice that high 2 and low 1´ resistance states are stable in zero magnetic field at a temperature of 94 K.[147] (c) Heterostructure and test setup schematics. (d) Spin asymmetry of Sample A after cooling in 700 mT without a bias and with a bias of –400 V, respectively.[147]

Direct electric field control of magnetization without multiferroics or magnetoelasticity may also induce magnetoelectric coupling. Grutter *et al.* demonstrated an electric field dependence of the emergent ferromagnetic layer at CaRuO$_3$/CaMnO$_3$ interfaces [Fig. 13(c)].[147] The application of a −400 V bias voltage on CaMnO$_3$/CaRuO$_3$ heterostructures induced an increase of the magnetization by a factor of 2.86 ± 0.4 [Fig. 13(d)], ascribing to the driving conduction electrons from the CaRuO$_3$ across the interface into the CaMnO$_3$, which enhanced the interfacial double exchange interaction.

The current information storage devices are based on two different aspects of solid state materials, spin and charge. Generally, spin and charge of electrons are employed separately. In magnetic recording and magnetoresistive random access memories (MRAM), a magnetic field (*H*) or a high-density current is used to write or read the information stored on the magnetization. Differently, the recording of two logic states (high resistive state HRS and low resistive state LRS) in resistive random access memories (RRAM) is on the basis of resistance switching modulated by electrical stimuli. The quest for higher data density in information storage is motivating researches to study how to manipulate magnetization without the need of cumbersome *H* and achieve a new paradigm where spin and charge act on each other. Thus with a simple RRAM architecture, electrical control of magnetism in RRAM with the medium of diluted magnetic oxides,[148,149] antiferromagnet,[150,151] manganites,[152,153] and ferrimagnets[154] has attracted extensive attention. The orderly migration of oxygen vacancies ($V_O$) is considered to the origin of the resistive switching (RS) behavior in these systems. Considering such structures not only show stable bipolar RS characteristics, but also exhibit magnetic modulation with the alternation of set and reset process, it is expected to obtain four logic states by encoding information in both RS and magnetic manipulation and even to achieve the dream of storing information magnetically and switching it electrically. Interestingly, magnetoresistance measurements indicate spin transport through an electrochemically formed copper nano-filament in LRS, in contrast to the disappearance of in HRS at the same temperature.[155]

# 5. Summary and outlook

In this review, we aim at illustrating the most recent progress in the electrical control of magnetism in oxides, whereas it is still very hard to cover all aspects, since this topic is continuing to develop so rapidly. The control of magnetism by ferroelectric switching in multiferroic heterostructures has focused mainly on strain, charge, and exchange bias couplings to achieve an effective interaction between the magnetic and ferroelectric components of the composite system. It gives rise to magnetoelectric couplings which are much larger than those typical of intrinsic multiferroics and functionalities that can in principle be optimized for device applications. The ionic liquid or EDL gating is a novel route to producing a dramatically large electric field using a voltage of only a few volts. The high permittivity and electric field in EDL gating enhance its ability on the manipulation of magnetism, reflecting in the remarkable ferromagnetic-antiferromagnetic (paramagnetic) phase transition under electric field. The electric field applied on the ordinary dielectric oxide could be used to control of the exotic phenomenon originated from the charge transfer and orbital reconstruction at the interface between dissimilar correlated oxides. The electrical control of magnetism in oxides has become a popular topic of ever-increasing interest in last decade, however, it is still in the quite infant stage, and thus there remain many open questions, both in mechanism and practical application:

(1) The mechanisms of the ME coupling remain to be clarified in details and extended. The nature of the strain-, charge-, and exchange bias-mediated ME coupling is somehow overlapped as we have mentioned in section 2. The mutual interaction of different ME coupling mechanisms is an important issue with reducing thickness of the films. Do these mechanisms work together or compete with each other? In addition, the strain, charge, and exchange bias respectively refers the lattice, charge, and spin degree of freedom in correlated oxide systems, whether the missing member orbital could mediate electric field and magnetic modulation in some form?[156]

(2) What is the role of oxygen vacancies in electrical control of magnetism? The

existence of oxygen vacancies with positive charges is inevitable in oxides, thus the formation, annihilation, and migration of oxygen vacancies under electric field are important to the electrical control of magnetism. The role of oxygen vacancies in both EDL and oxide gating needs to be demonstrated in the future researches.

(3) How to optimize and enhance the effect of electric field on magnetism? In the FET with oxide insulator, a thick oxide insulator is needed for preventing the leakage current but weakens the amplitude of electric field applied on the magnetic layer. Thus only the materials near the FM-AFM transition points behave significant dependences on the electric field in the carrier-mediated cases. On the other hand, the effective thickness in metallic FM material is only a few nanometers, which limits the manipulation of magnetic performance.

(4) Much of the works carried out thus far in this field pertain to quasi-static phenomena. One exception is the electric field control of magnetic resonances in the GHz range, where a modulation of the spin-wave spectra has been demonstrated. However, the dynamic magnetoelectric response still needs to be addressed in more detail.

(5) The practical application of electrical control of magnetism in oxides calls for more endeavors: i) the device design and integration for electrical control of magnetism; ii) the enhancement of operated temperature upon room temperature; iii) the epitaxial growth and precise control of multiferroic hetero-films with desired structures by sputtering; iv) the integration of ionic liquid in traditional semiconductor industry.